# Approximate longitudinal space charge impedances of a round beam between parallel plates and inside a rectangular chamber

Yingjie Li[a,*], Lanfa Wang[b]

[a]*Department of Physics, Michigan State University, East Lansing, MI 48824, USA*
[b]*SLAC National Accelerator Laboratory, Menlo Park, CA 94025, USA*

**Abstract**

This paper presents the approximate analytical solutions to the longitudinal space charge (LSC) impedances of a round beam with uniform transverse distribution and sinusoidal line density modulations under two boundary conditions: (a) between parallel plates (b) inside a rectangular chamber, respectively. When the ratio of beam diameter to chamber height is small, the image charge fields of the round beam can be approximated by those of a line charge, the approximate analytical LSC impedances can be obtained by image method. The derived theoretical LSC impedances are valid at any perturbation wavelength and are consistent well with the numerical simulation results in a large range of ratios of beam diameters to chamber heights.

## 1. Introduction

The longitudinal space charge (LSC) effect plays an important role in the microwave instability of low energy beam with high intensity near or above transition [1-3]. It is also one of the important reasons causing the micro-bunching instability for free-electron lasers (FELs) [4]. An accurate calculation of the LSC fields and impedances is helpful to explain the beam behavior and predict the growth rates of the beam instability. Both the direct self-fields and the image charge fields due to the conducting chamber wall should be taken into account in the calculation of the LSC fields and impedances. The image charges may reduce the LSC fields inside the beam and the associated LSC impedances compared with a beam in free space, this is just the so-called shielding effect of the vacuum chamber. Various space charge field models have been investigated in existing literatures. For example, a round beam in free space [5-8], a round beam inside a round chamber [3, 6, 7, 9, 10], a round beam inside an elliptic chamber [11], a uniformly charged line between two parallel plates [12], a uniformly charged round beam between two parallel plates [13], a uniformly charged round beam inside a rectangular chamber [14], a rectangular beam inside a rectangular chamber [15-17], a rectangular beam between two parallel plates [17, 18], etc.

Some popular methods are used to calculate the analytical LSC fields (a) Faraday's law and rectangular integration loop [14, 19]. This method is only valid when the perturbation wavelengths $\lambda$ of the longitudinal charge density are much longer than the transverse chamber dimensions (long-wavelength limits). When the charge density modulation wavelengths are small, the electric fields at the off-axis field points have both normal and skew components with respect to the beam axis, the three-dimensional (3D) effects of the electric fields become important making this method invalid. (b) Direct integration methods. Usually the direct integration methods are only applicable to the field models with simple charge distributions in free space. Some literatures use this method to calculate the LSC fields assuming the gradient of the charge density $d\Lambda/dz$ is independent of the longitudinal coordinate $z$ and is put outside of the integral over $z$ (e.g., Refs. [18, 19]). In fact, this assumption is invalid for a beam with short-wavelength density modulations (e.g., $\Lambda(z) = \Lambda_k \cos(kz)$, where $k=2\pi/\lambda$). Thus the results are only valid in the long-wavelength limits too. (c) Separation of variables. In some special cases, the exact analytical 3D space charge fields of a beam with sinusoidal longitudinal charge density modulations can be solved by the method of separation of variables, such as a round beam in free space and inside a round chamber [3, 5, 9], a rectangular beam inside a rectangular chamber [17], etc. The space charge fields solved by this method are exact and valid in the whole spectrum of perturbation wavelengths. But this method is critical to the configurations of the cross-sectional geometry of the beam-chamber system, hence it is not applicable to all the field models. (d) Image method. According to the superposition theorem of the electric fields, the space charge field of a beam is equal to the sum of the direct self-field in free space (open boundary) and its image field. If these fields can be calculated separately, it is easy to obtain the total LSC field and impedance.

---

\* Corresponding author.
*Email address*: liyingji@msu.edu (Y. Li).

In many accelerators, the cross-sections of the beam and vacuum chamber are approximately round and rectangular, respectively. Usually the rectangular vacuum chamber has a large aspect ratio, hence it can also be approximated by a pair of infinitely large parallel plates. For example, this assumption is used in the 3D particle-in-cell (PIC) simulation code CYCO [20]. In order to study the microwave and micro-bunching instability conveniently, beam physicists are interested in obtaining analytical expressions for the LSC impedances of a round beam inside a straight rectangular chamber or between parallel plates. In addition, it is highly desirable that the analytical expressions are accurate enough at any perturbation wavelengths. Though there are some existing literatures discussing the longitudinal coupling impedances of rectangular chambers and parallel plates, these models either are only valid in the long-wavelength limits [14], or are not for a round beam model (e.g., single particle [21], line charge [22], vertical ribbon beam [23], etc), or use numerical methods (e.g., Refs. [24-25], etc). To our knowledge, at present, there are no analytical expressions available in modern publications for the LSC impedances of a round beam inside a straight rectangular chamber or between parallel plates which are valid at any perturbation wavelengths.

This paper provides approximate analytical LSC fields and impedances of an infinitely long, straight and round beam with uniform transverse density between parallel plates and inside a rectangular chamber. The solutions are valid at any perturbation wavelengths. We know, if a round beam has longitudinal charge density modulations, its 3D space charge fields inside a rectangular chamber or between parallel plates cannot be solved directly and exactly by the separation of variables technique. Since the transverse dimensions of almost all the beam chambers are much larger than the transverse beam size, the image charge fields of a round beam can be approximated by those of a line charge. Then the approximate LSC fields and impedances of the two models in discussion can be calculated by image methods. Firstly, using the method of separation of variables, the LSC fields of a line charge with density modulations between two parallel plates and inside a rectangular chamber can be solved, respectively. Secondly, the space charge fields of a line charge and a round beam with longitudinal charge density modulations in free space are also calculated, respectively. Finally, the LSC fields and impedances of the round beam between parallel plates and inside the rectangular chamber can be approximated by the superposition theorem of the electric fields, respectively.

This paper is organized as follows. Section 2 provides the analytical solutions to the LSC fields of four component field models that are necessary for the derivations of the LSC impedances in discussion. Section 3 derives the analytical LSC impedances of a round beam between parallel plates and inside a rectangular chamber. Section 4 compares the theoretical and simulated LSC impedances with predictions of other existing impedance models in both the short-wavelength and long-wavelength limits.

## 2. Space charge field models

In order to obtain the approximate analytical LSC impedances of a round beam with planar and rectangular boundary conditions, first, we need to know the LSC fields $E_z$ of the following four component field models: (a) A round beam in free space, $E_{z,round,fs}$. (b) A line charge in free space, $E_{z,line,fs}$. (c) A line charge between two parallel plates, $E_{z,line,pp}$. (d) A line charge inside a rectangular chamber $E_{z,line,rect}$. For a *round* beam between a pair of parallel plates, when the separation between the plates is much larger than the beam diameter, its image LSC fields can be approximated by those of a line charge between the parallel plates as $E_{z,round,pp}^{image} \approx E_{z,line,pp}^{image} = E_{z,line,pp} - E_{z,line,fs}$, its total LSC fields are approximately equal to $E_{z,round,pp} = E_{z,round,fs} + E_{z,round,pp}^{image} \approx E_{z,round,fs} + E_{z,line,pp}^{image} = E_{z,round,fs} + E_{z,line,pp} - E_{z,line,fs}$; similarly, for a *round* beam inside a *rectangular* chamber, when the full chamber height is much larger than the beam diameter, its image LSC fields $E_{z,round,rect}^{image}$ and total LSC fields $E_{z,round,rect}$ can be approximated as $E_{z,round,rect}^{image} \approx E_{z,line,rect} - E_{z,line,fs}$ and $E_{z,round,rect} \approx E_{z,round,fs} + E_{z,line,rect} - E_{z,line,fs}$ respectively. Next, we will derive the LSC fields of the four component field models listed in (a)-(d).

*2.1. A round beam in free space*

In the *lab* frame, assume there is an infinitely long, transversely uniform round beam of radius $r_0$ with sinusoidal line density $\Lambda$ and beam intensity modulations $I$ of

$$\Lambda(z,t) = \Lambda_k \exp[i(kz - \omega t)], \text{ and } I(z,t) = I_k \exp[i(kz - \omega t)], \tag{1}$$

respectively, where $\Lambda_k$ and $I_k$ are the amplitudes, $I_k = \Lambda_k \beta c$, $\beta$ is the relativistic speed of the beam, $c$ is the speed of light, $\omega$ is the angular frequency of the perturbations, $k$ is the wavenumber of the line charge density modulations. According to Ref. [5], its LSC fields in the *lab* frame are

$$E_{z,round,fs}(r,z,t) = -\frac{1}{\pi \varepsilon_0 r_0^2 \bar{k}^2 \gamma^2} \frac{\partial \Lambda(z,t)}{\partial z} [1 - K_1(\bar{k}r_0) I_0(\bar{k}r)]. \tag{2}$$

where $\varepsilon_0 = 8.85 \times 10^{-12}$ F m$^{-1}$ is the permittivity in free space, $\bar{k} = k/\gamma$, $\gamma$ is the relativistic factor, $I_0(x)$ and $K_1(x)$ are the modified Bessel functions of the first and second kinds, respectively.

*2.2. A line charge in free space*

In the *lab* frame, assume there is an infinitely long line charge in free space with sinusoidal line charge density and beam intensity modulations described in Eq. (1). First we can calculate its potentials and fields in the *rest* frame of the beam, and then convert them into the *lab* frame by Lorentz transformation. In the *rest* frame of the beam, the line charge density is

$$\Lambda'(z') = \Lambda'_k \cos(k'z'), \tag{3}$$

where the parameters with primes in this paper stand for those in the *rest* frame. The electrostatic potentials can be calculated easily in cylindrical coordinate system by direct integration as

$$\varphi'_{line,fs}(r',z') = \frac{\Lambda'_k}{4\pi\varepsilon_0} \int_{-\infty}^{\infty} \frac{\cos(k'\bar{z}')}{[(\bar{z}'-z')^2 + r'^2]^{\frac{1}{2}}} d\bar{z}' = \frac{\Lambda'(z')}{2\pi\varepsilon_0} K_0(k'r'). \tag{4}$$

The LSC fields in the *rest* frame are

$$E'_{z,line,fs}(r',z') = -\frac{1}{2\pi\varepsilon_0} \frac{d\Lambda'(z')}{dz'} K_0(k'r'). \tag{5}$$

In the *lab* frame, according to the theory of relativity, we have

$$E'_z = E_z, \tag{6}$$

$$r' = r, \tag{7}$$

$$z' = \gamma(z - \beta ct), \tag{8}$$

$$\Lambda'_k = \Lambda_k / \gamma, \tag{9}$$

$$k' = \frac{k}{\gamma} = \bar{k}, \tag{10}$$

$$\frac{d\Lambda'(z')}{dz'} = -k'\Lambda'_k \sin(k'z') = -\frac{k\Lambda_k}{\gamma^2}\sin(kz-\omega t). \tag{11}$$

If we choose exponential representation as used in Eq. (1), then Eq. (11) can also be expressed as

$$\frac{d\Lambda'(z')}{dz'} = \frac{1}{\gamma^2}\frac{\partial \Lambda(z,t)}{\partial z}. \tag{12}$$

From Eqs. (5)-(12), the LSC fields in the *lab* frame become

$$E_{z,line,fs}(r,z,t) = -\frac{1}{2\pi\varepsilon_0\gamma^2}\frac{\partial \Lambda(z,t)}{\partial z}K_0(\bar{k}r). \tag{13}$$

*2.3. A line charge between two parallel plates*

The schematic view of an infinitely long line charge between two infinitely large, perfectly conducting parallel plates is shown in Fig. 1. Its sinusoidal line charge density and beam intensity modulations are described by Eq. (1). Assume the two plates are separated by a distance *H*, the line charge is parallel to the plates and its distance to the lower plate is $\Gamma$, the potentials on the two plates are all 0. Though Ref. [22] provided solutions to the LSC fields and impedances of a line charge between parallel plates and inside a rectangular chamber, the field potential is solved by 2D Green function neglecting the 3D effects caused by the line density modulations. Hence the results are only valid in the long-wavelength limits. Ref. [26] solved the 2D electrostatic potentials of a uniform line charge between two parallel plates using the method of separation of variables. We can use the same method and similar procedures to solve the 3D fields of our model. We choose the Cartesian coordinate *xoy* with *o* as the origin. Assume in the *rest* frame of the beam, the basic harmonic component of the space charge potential can be written in the form

$$\varphi'_{line,pp}(x',y',z') = X(x')Y(y')\cos(k'z'), \tag{14}$$

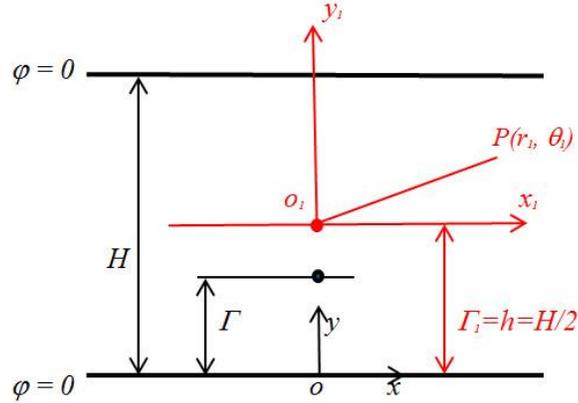

Fig. 1. A line charge with sinusoidal density modulations between parallel plates.

which satisfies the Laplace equation

$$\frac{\partial^2 \varphi'_{line,pp}}{\partial x'^2} + \frac{\partial^2 \varphi'_{line,pp}}{\partial y'^2} + \frac{\partial^2 \varphi'_{line,pp}}{\partial z'^2} = 0. \tag{15}$$

Plugging Eq. (14) into Eq. (15) resulting in

$$\frac{1}{X}\frac{d^2X}{dx'^2} + \frac{1}{Y}\frac{d^2Y}{dy'^2} = k'^2. \tag{16}$$

Considering the boundary conditions $\varphi'_{line,pp}(y'=0) = \varphi'_{line,pp}(y'=H) = \varphi'_{line,pp}(x'=\pm\infty) = 0$, we can choose

$$\frac{1}{X}\frac{d^2X}{dx'^2} = k'^2 + \alpha^2, \quad \frac{1}{Y}\frac{d^2Y}{dy'^2} = -\alpha^2, \tag{17}$$

where $\alpha > 0$. Then the solutions to Eq. (17) can be written as

$$X(x') = A'_1 e^{\sqrt{k'^2+\alpha^2}\,x'} + A'_2 e^{-\sqrt{k'^2+\alpha^2}\,x'}, \tag{18}$$

$$Y(y') = B'_1 \sin(\alpha y') + B'_2 \cos(\alpha y'). \tag{19}$$

The boundary conditions $\varphi'_{line,pp}(y'=0) = \varphi'_{line,pp}(y'=H) = 0$ give $B'_2 = 0$, $\alpha = n\pi/H$, then $Y(y') \sim \sin(n\pi y'/H)$. Because at $x'=0$, there is a line charge which produces singularity, we should calculate the electrostatic potentials $\varphi'_{line,pp,+}$ for $x' > 0$ and $\varphi'_{line,pp,-}$ for $x' < 0$ separately. In Eq. (18), when $x' \to +\infty$, $\varphi'_{line,pp,+} \to 0$, then the coefficient $A'_{1+} = 0$; when $x' \to -\infty$, $\varphi'_{line,pp,-} \to 0$, then the coefficient $A'_{2-} = 0$. The solutions of $X$ can be written as

$$X_{\pm}(x') = A'_{\pm} e^{\mp\sqrt{k'^2+\alpha^2}\,x'}. \tag{20}$$

where '+' and '−' stand for x>0 and x<0, respectively.

The potentials including all harmonic components can be expressed as

$$\varphi'_{line,pp,\pm} = \sum_{n=1}^{\infty} C'_{n\pm} e^{\mp\sqrt{k'^2+\left(\frac{n\pi}{H}\right)^2}\,x'} \sin(\frac{n\pi}{H} y') \cos(k'z'), \tag{21}$$

where $C'_{n+}$ and $C'_{n-}$ are the coefficients to be determined by the boundary conditions for $x>0$ and $x<0$, respectively. At $x'=0$, $y'\neq\Gamma$, $\varphi'_{line,pp,+} = \varphi'_{line,pp,-}$ which gives $C'_{n+} = C'_{n-} = C'_n$. If the line charge is rewritten in the form of surface charge density

$$\sigma' = \Lambda'(z')\delta(y' - \Gamma), \tag{22}$$

where $\delta(x)$ is the Dirac Delta function, then on the plane $x'=0$, the boundary condition $D_{2n}' - D_{1n}' = \sigma'$ gives

$$\varepsilon_0\left(\frac{\partial \varphi'_{line,pp,-}}{\partial x'} - \frac{\partial \varphi'_{line,pp,+}}{\partial x'}\right)\bigg|_{x'=0} = \Lambda'_k \cos(k'z')\delta(y' - \Gamma). \tag{23}$$

Eqs. (21) and (23) yield

$$\sum_{n=1}^{\infty} 2C'_n \sqrt{k'^2 + \left(\frac{n\pi}{H}\right)^2} \sin(\frac{n\pi}{H} y') = \frac{\Lambda'_k}{\varepsilon_0} \delta(y' - \Gamma). \tag{24}$$

Multiplying the two sides of Eq. (24) by $sin(n\pi y'/H)$ and integrating $y'$ from 0 to $H$ gives the coefficient $C'_n$

$$C'_n = \frac{\Lambda'_k}{\varepsilon_0 H \sqrt{k'^2 + \left(\frac{n\pi}{H}\right)^2}} \sin(\frac{n\pi}{H} \Gamma). \tag{25}$$

Then the potentials in Eq. (21) can be expressed as

$$\varphi'_{line,pp}(x', y', z') = \frac{\Lambda'(z')}{\varepsilon_0 H} \sum_{n=1}^{\infty} \frac{1}{\sqrt{k'^2 + \left(\frac{n\pi}{H}\right)^2}} \sin(\frac{n\pi}{H} \Gamma) e^{-\sqrt{k'^2 + \left(\frac{n\pi}{H}\right)^2}|x'|} \sin(\frac{n\pi}{H} y'). \tag{26}$$

Let's consider a special case of $\Gamma = \Gamma_1 = h = H/2$, i.e., the line charge is on the median plane of the two plates as shown in Fig. 1. If we choose a new coordinate system $x_1 o_1 y_1$ with $O_1$ as the origin (see Fig. 1), according to $x' = x_1'$, $y' = y_1' + h$, the potentials in the *rest* frame of the beam become

$$\varphi'_{line,pp}(x_1', y_1', z') = \frac{\Lambda'(z')}{2\varepsilon_0 h} \sum_{n=1}^{\infty} \frac{1}{\sqrt{k'^2 + \left(\frac{n\pi}{2h}\right)^2}} \sin(\frac{n\pi}{2}) e^{-\sqrt{k'^2 + \left(\frac{n\pi}{2h}\right)^2}|x_1'|} \sin[\frac{n\pi}{2h}(y_1' + h)]. \tag{27}$$

If we use cylindrical coordinate system, $x_1' = r'\cos(\theta')$, $y_1' = r'\sin(\theta')$, Eq. (27) becomes

$$\varphi'_{line,pp}(r', \theta', z') = \frac{\Lambda'(z')}{2\varepsilon_0 h} \sum_{n=1}^{\infty} \frac{1}{\sqrt{k'^2 + \left(\frac{n\pi}{2h}\right)^2}} \sin(\frac{n\pi}{2}) e^{-\sqrt{k'^2 + \left(\frac{n\pi}{2h}\right)^2} r'|\cos(\theta')|} \sin[\frac{n\pi}{2h}(r'\sin(\theta') + h)]. \tag{28}$$

Using the Lorentz transformation of Eqs. (6)-(12), and $\theta' = \theta$, the LSC fields in the *lab* frame become

$$E_{z,line,pp}(r, \theta, z, t) = -\frac{1}{2\varepsilon_0 h \gamma^2} \frac{\partial \Lambda(z,t)}{\partial z} \sum_{n=1}^{\infty} \frac{1}{\sqrt{\bar{k}^2 + \left(\frac{n\pi}{2h}\right)^2}} \sin(\frac{n\pi}{2}) e^{-\sqrt{\bar{k}^2 + \left(\frac{n\pi}{2h}\right)^2} r|\cos(\theta)|} \sin[\frac{n\pi}{2h}(r\sin(\theta) + h)]. \tag{29}$$

*2.4. A line charge inside a rectangular chamber*

In the *lab* frame, assume there is an infinitely long line charge inside a rectangular chamber, the sinusoidal line charge density and beam intensity modulations are described in Eq. (1). The full chamber width and height are $W=2w$ and $H=2h$, respectively. Ref. [17] derived the potentials of an infinitely long beam with rectangular cross-section and uniform transverse charge density inside a rectangular chamber. The full beam width and height are $2a$ and $2b$, respectively. In the *rest* frame of beam, in the charge-free region inside the chamber ($b \leq |y'| \leq h$), the potentials are

$$\varphi'_{rect,rect}(x',y',z') = \frac{\Lambda'_k \cos(k'z')}{4\varepsilon_0 bw} \sum_{n=1}^{\infty} \frac{g'_n}{v_n'^2} \frac{\sinh(v'_n b)}{\cosh(v'_n h)} \sin[\eta_n(x'+w)]\sinh[v'_n(h-|y'|)], \tag{30}$$

where
$$g'_n = g_n = \frac{2}{a\eta_n} \sin(\eta_n w)\sin(\eta_n a), \tag{31}$$

and $\eta_n = n\pi/2w$, $v_n'^2 = \eta_n^2 + k'^2$, $n=1, 2, 3,\ldots$. In the limiting case of $a=b=0$, the rectangular beam shrinks to a line charge. Because $a=0$, $g_n' = g_n = 2\sin(\eta_n w)$ and $b=0$, $\sinh(v_n' b)/b = v_n'$, then Eq. (30) becomes

$$\varphi'_{line,rect}(x',y',z') = \frac{\Lambda'(z')}{2\varepsilon_0 w} \sum_{n=1}^{\infty} \frac{\sin(\eta_n w)}{v'_n \cosh(v'_n h)} \sin[\eta_n(x'+w)]\sinh[v'_n(h-|y'|)]. \tag{32}$$

Using the Lorentz transformation of $x'=x$, $y'=y$, and Eqs. (6), (8)-(12), the LSC fields in the *lab* frame become

$$E_{z,line,rect}(x,y,z,t) = -\frac{1}{2\varepsilon_0 w \gamma^2} \frac{\partial \Lambda(z,t)}{\partial z} \sum_{n=1}^{\infty} \frac{\sin(\eta_n w)}{v_n \cosh(v_n h)} \sin[\eta_n(x+w)]\sinh[v_n(h-|y|)], \tag{33}$$

where $v_n^2 = v_n'^2 = \eta_n^2 + \bar{k}^2 = \eta_n^2 + k^2/\gamma^2$.

## 3. LSC impedances of a round beam between parallel plates and inside a rectangular chamber

There exist several different definitions for the longitudinal impedances [7, 22, 27, etc]. For example, in Ref. [27], if a beam has sinusoidal current as

$$I(s,t) = I_k \exp[i(ks - \omega t)], \tag{34}$$

where $s$ is the path length traveled by a particle with respect to a fixed position of accelerator ring, for an accelerator section with length $L$, the longitudinal wake field potential $V(s, t)$ (it is equal to the integrated longitudinal electric field, or energy loss of a unit charge over the accelerator section) is

$$V(s,t) = -I(s,t) Z_0^{\parallel}(\omega), \tag{35}$$

where $Z_0^{\parallel}(\omega)$ is the longitudinal impedance for $m=0$ mode at frequency $\omega$. For the space charge effects in an accelerator ring, set $L=2\pi R$, where $R$ is the average ring radius, $Z_0^{\parallel}(\omega)$ can be calculated from Fourier transform of the longitudinal wake function $W_0'(\xi)$ as

$$Z_0^{\parallel}(\omega) = \int_{-\infty}^{\infty} \frac{d\xi}{\beta c} e^{-i\omega\xi/\beta c} W_0'(\xi), \tag{36}$$

where $\xi = s - s'$ is the distance between the field point $s$ and source point $s'$. Specifically, $W_0'(\xi)=0$ for $\xi > 0$. For the convenience of instability analysis for a given perturbation wavenumber $k$, some beam physicists prefer to express the LSC impedances directly as a function of $k$ instead of $\omega$ (for example, Refs. [2, 4, 6, 7, 8, 17, 22]). If we choose the bunch center as reference point for the longitudinal coordinate, Eqs. (34) and (35) can be rewritten as function of $z$ instead of $s$, this parameter replacement should not affect the expression of LSC impedance which is a property in frequency domain. For a given beam current as described in Eq. (1), if we know the induced LSC field $E_z(z, t)$ and LSC wake potential $V(z, t)$, we can calculate LSC impedance as a function of $k$ conveniently by Eq. (35) without resorting to the LSC wake function and its Fourier transform in Eq. (36) (see Ref. [17]). According to Eqs. (1) and (35), the *average* longitudinal wake potential (or energy loss per turn of a unit charge) in a circular accelerator due to the LSC fields is

$$<V(z,t)> = -<E_z>C_0 = Z_0^{\|}(k) I_k \exp[i(kz-\omega t)], \tag{37}$$

where $<E_z>$ is the LSC fields averaged over the cross-section of the beam and can be calculated using the formula

$$<f(r,\theta)> = \frac{1}{\pi r_0^2} \int_0^{2\pi} d\theta \int_0^{r_0} f(r,\theta) r dr. \tag{38}$$

(a) For a round beam midway between parallel plates, the *average* LSC impedance can be calculated by Eqs. (2), (13), (29), and (37) with $E_z = E_{z,round,pp} \approx E_{z,round,fs} + E_{z,line,pp} - E_{z,line,fs}$ as

$$Z_{0,round,pp}^{\|}(\bar{k}) = i\frac{Z_0 C_0}{2\beta\gamma h}\chi_{line,pp}(\bar{k}) + i\frac{Z_0 C_0}{\pi\beta\gamma r_0} K_1(\bar{k}r_0)[1-\frac{2 I_1(\bar{k}r_0)}{\bar{k}r_0}], \tag{39}$$

where

$$\chi_{line,pp}(\bar{k}) = \sum_{n=1}^{\infty} \frac{1}{\sqrt{1+\left(\frac{n\pi}{2\bar{k}h}\right)^2}} \sin(\frac{n\pi}{2}) < e^{\sqrt{1+\left(\frac{n\pi}{2\bar{k}h}\right)^2}\bar{k}r|\cos(\theta)|} \sin[\frac{n\pi}{2h}(r\sin(\theta)+h)] >. \tag{40}$$

(b) For a round beam inside and coaxial with a rectangular chamber, the *average* LSC impedance can be calculated by Eqs. (2), (13), (33), and (37) with $E_z = E_{z,round,rect} \approx E_{z,round,fs} + E_{z,line,rect} - E_{z,line,fs}$ as

$$Z_{0,round,rect}^{\|}(\bar{k}) = i\frac{Z_0 C_0}{2\beta\gamma w}\chi_{line,rect}(\bar{k}) + i\frac{Z_0 C_0}{\pi\beta\gamma r_0} K_1(\bar{k}r_0)[1-\frac{2 I_1(\bar{k}r_0)}{\bar{k}r_0}], \tag{41}$$

where

$$\chi_{line,rect}(\bar{k}) = \sum_{n=1}^{\infty} \frac{\bar{k}\sin(\eta_n w)}{v_n \cosh(v_n h)} <\sin[\eta_n(r\cos(\theta)+w)]\sinh[v_n(h-r|\sin(\theta)|)]>. \tag{42}$$

In the derivations of Eqs. (39) and (41), two identities of integrals $\int_0^{\xi} x K_0(x) dx = 1 - \xi K_1(\xi)$ and $\int_0^{\xi} x I_0(x) dx = \xi K_1(\xi)$ are used. Note the first terms on the right hand side of Eqs. (39) and (41) are contributed from the *average* LSC fields of a line charge midway between parallel plates and inside a rectangular chamber, respectively; the second terms are contributed from the differences of the *average* LSC fields within beam radius $r_0$ between a round beam and a line charge in free space. Eqs. (40) and (42) can be evaluated by truncating the infinite series to a finite number of terms, as long as the sum converges well.

## 4. Case study and comparisons of LSC impedances

In this section, we will calculate the approximate LSC impedances of SIR beam using the formulae derived for the two field models in Section 3. For the purpose of comparisons, first we would like to summarize some LSC impedance formulae which are often used in modern beam physics literatures.

*4.1. LSC impedance models for a round beam*

*4.1.1. A round beam inside a round chamber*

For a round beam with radius $r_0$ and uniform transverse distribution centered inside a round chamber with inner chamber wall radius $r_w$, the LSC impedance is

$$Z^{\|}_{0,round,round}(\bar{k}) = i\frac{2RZ_0}{\beta\gamma\bar{k}r_0^2}\{1 - \frac{f_1}{I_0(\bar{k}r_w)}[K_1(\bar{k}r_0)I_0(\bar{k}r_w) + K_0(\bar{k}r_w)I_1(\bar{k}r_0)]\}. \tag{43}$$

where $f_1 = \bar{k}r_0$ for *on-axis* impedance [6, 7] and $f_1 = 2I_1(\bar{k}r_0)$ for the *average* one [17], respectively. The identity of $<I_0(\bar{k}r)> = 2I_1(\bar{k}r_0)/\bar{k}r_0$ is used for the case of *average* impedance.

(a) In the long-wavelength limits

The total LSC impedance of a uniform disk beam with radius $r_0$ inside a round chamber with radius $r_w$ in the *long-wavelength* limits is [7, 27]

$$Z^{\|,LW}_{0,round,round}(\bar{k}) = i\frac{\bar{k}RZ_0}{\beta\gamma}(f_2 + \ln\frac{r_w}{r_0}), \tag{44}$$

where $f_2 = 1/2$ for the *on-axis* impedance and $f_2 = 1/4$ for the *average* one, respectively.

(b) In the short-wavelength limits

If $r_w \gg r_0$, the image charge effects of the chamber wall can be neglected in the short-wavelength limits, the LSC impedance of a round beam is approximately equal to that in free space. Refs. [4, 5] give the *on-axis* LSC impedance of a round beam in the short-wavelength limits as

$$Z^{\|,axis,SW}_{0,round,round}(\bar{k}) = Z^{\|,axis}_{0,round,fs}(\bar{k}) = i\frac{2RZ_0}{\beta\gamma\bar{k}r_0^2}[1 - \bar{k}r_0 K_1(\bar{k}r_0)]. \tag{45}$$

The LSC impedances in Eq. (45) are derived from the *on-axis* LSC fields of the 1D space charge field model. While Ref. [7] pointed out that the 1D field model does not hold any more for $\lambda < 4\pi r_0/\gamma$ or $kr_0/\gamma > 0.5$. In addition, the *off-axis* LSC fields always decrease from the beam axis $r=0$ to the beam edge $r=r_0$. Ref. [28] studied these 3D space charge effects analytically and made a conclusion that, if the LSC fields were averaged over the beam cross-section, the 1D and 3D field models predict almost the identical LSC fields. The *average* LSC impedance is given in Refs. [6, 8] as

$$Z^{\|}_{0,round,fs}(\bar{k}) = i\frac{2RZ_0}{\beta\gamma\bar{k}r_0^2}[1 - 2I_1(\bar{k}r_0)K_1(\bar{k}r_0)]. \tag{46}$$

*4.1.2. A round beam inside a rectangular chamber in the long-wavelength limits*

Let's assume an infinitely long, transversely uniform round beam with radius $r_0$ is inside and coaxial with a rectangular chamber. The full chamber width and height are $W=2w$ and $H=2h$, respectively. Then according to Eq. (23) of Ref. [14], the LSC impedance of an accelerator ring in the *long-wavelength* limits is:

$$Z^{\|,LW}_{0,round,rect}(\bar{k}) = i\frac{\bar{k}RZ_0}{\beta\gamma}\{f_3 + \ln[\frac{4h}{\pi r_0}\tanh(\frac{\pi w}{2h})]\}, \tag{47}$$

where $f_3 = 1/2$ for the *on-axis* impedance and $f_3 = 1/4$ for the *average* one, respectively.

*4.1.3. A round beam between parallel plates in the long-wavelength limits*

In the limiting case of $W \to \infty$, the rectangular chamber becomes a pair of parallel plates, according to Eq. (47), Eqs. (A6) and (A7) in Appendix of Ref. [13], its LSC impedance becomes

$$Z_{0,round,pp}^{\|,LW}(\bar{k}) = i \frac{\bar{k} R Z_0}{\beta \gamma}[f_4 + \ln(\frac{4h}{\pi r_0})], \tag{48}$$

where $f_4 = 1/2$ for the *on-axis* impedance and $f_4 = 1/4$ for the *average* one, respectively.

*4.2. LSC impedance models for a rectangular beam*

Ref. [17] derived the *average* LSC impedances of a rectangular beam between parallel plates and inside a rectangular chamber, respectively. Please refer to Ref. [17] for the detailed formulae.

*4.3. Case study and comparisons of LSC impedances*

As a case study, we can calculate the LSC impedances of a coasting $H_2^+$ ion beam in the Small Isochronous Ring (SIR) at Michigan State University (MSU) [2, 20]. The ring circumference is $C_0 = 6.58$ m, the kinetic energy of the beam is $E_k = 20$ keV ($\beta \approx 0.0046$, $\gamma \approx 1$), the beam radius is variable. The half width and half height of the rectangular vacuum chamber are $w = 5.7$ cm and $h = 2.4$ cm, respectively. Since $w \gg h$, the rectangular chamber can also be simplified as a pair of infinitely large horizontal parallel plates. The LSC impedances are calculated by both the theoretical and numerical methods. A general-purpose simulation code developed by us [17] based on the Finite Element Method (FEM) is used in the numerical simulations.

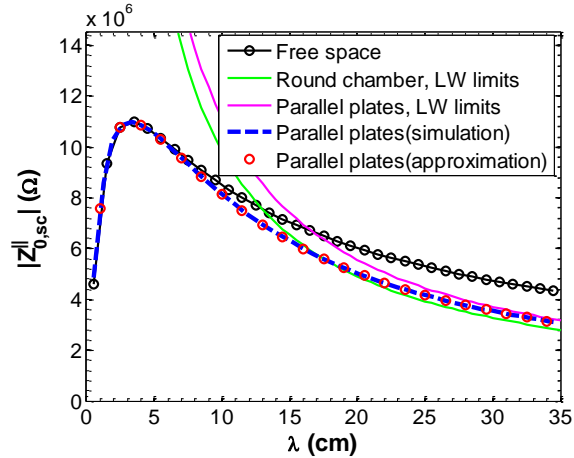

Fig. 2. Comparisons of the *average* LSC impedances of a *round* SIR beam with beam radius $r_0 = 0.5$ cm under different boundary conditions and in different wavelength limits. $\lambda$ is the perturbation wavelength, $|Z_{0,sc}^{\|}|$ is the modulus of LSC impedance. In the legend, 'Free space', 'Round chamber', and 'Parallel plates' are boundary conditions; 'LW limits' stands for the *long-wavelength* limits; '(approximation)' and '(simulation)' stand for the theoretical approximation and simulation (FEM) methods, respectively.

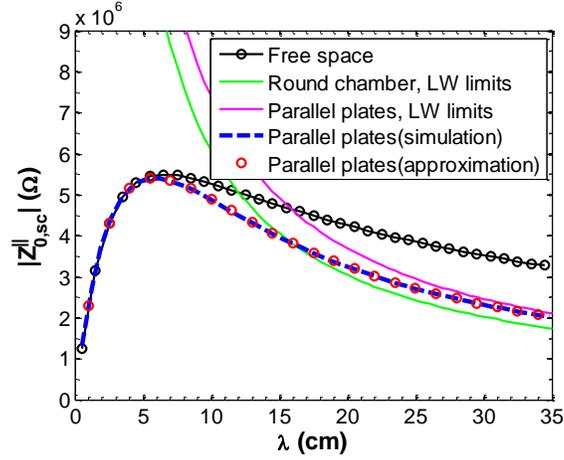

Fig. 3. Comparisons of the *average* LSC impedances of a *round* SIR beam with beam radius $r_0$=1.0 cm under different boundary conditions and in different wavelength limits.

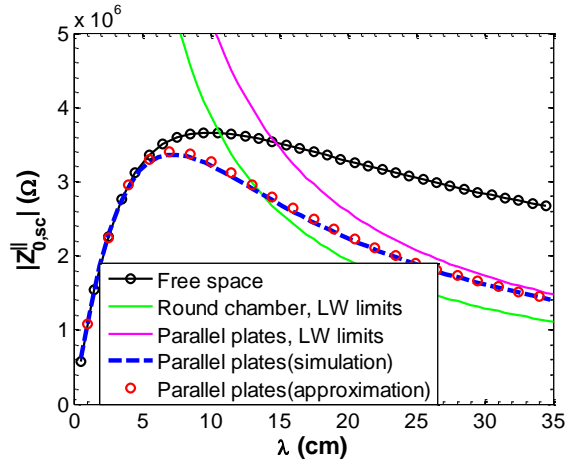

Fig. 4. Comparisons of the *average* LSC impedances of a *round* SIR beam with beam radius $r_0$=1.5 cm under different boundary conditions and in different wavelength limits.

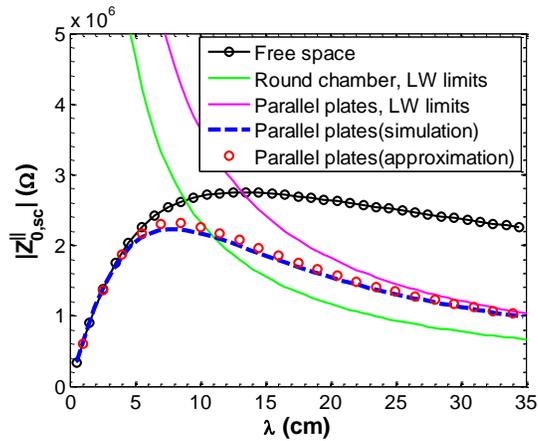

Fig. 5. Comparisons of the *average* LSC impedances of a *round* SIR beam with beam radius $r_0$=2.0 cm under different boundary conditions and in different wavelength limits.

Figs. 2-5 show the simulated (blue dashes) and theoretically approximated (Eqs. (39) and (40), red circles) *average* LSC impedances of a round SIR beam with radii $r_0$ =0.5 cm, 1.0 cm, 1.5 cm and 2.0 cm midway between the parallel plates with $h$=2.4 cm. For the purpose of comparisons, the *theoretical average* LSC impedances of the round beam predicted by three existing models are also plotted. (a) In free space (Eq. (46), black lines with circles). (b) Inside a round chamber with $r_w$=$h$=2.4 cm, in the long-wavelength limits (Eq. (44), green lines). (c) Between parallel plates with $h$=2.4 cm, in the long-wavelength limits (Eq. (48), magenta lines). For a small beam size, for instance $r_0$<1.0 cm, the theoretical approximations are consistent well with the simulations in all the wavelengths. A small discrepancy appears for large beam size case when the image charge effect becomes large, for instance $r_0$=2.0 cm. The long-wavelength model with a round chamber gives smaller impedance as expected because of the larger shielding effect compared with a pairs of parallel plates. The difference of the impedance between a round chamber and a pairs of parallel plates becomes larger when the beam size increases.

Figs. 6-9 show the simulated (blue dashes) and theoretically approximated (Eqs. (41) and (42), red circles) *average* LSC impedances of a round SIR beam with radii $r_0$ =0.5 cm, 1.0 cm, 1.5 cm and 2.0 cm inside and coaxial with a rectangular chamber with $w$=5.7 cm, $h$=2.4 cm. For the purpose of comparisons, the *theoretical average* LSC impedances predicted by three existing models are also plotted. (a) In free space (Eq. (46), black lines with circles). (b) Inside a round chamber with $r_w$=$h$=2.4 cm, in the long-wavelength limits (Eq. (44), green lines). (c) Inside a rectangular chamber with $w$=5.7 cm, $h$=2.4 cm, in the long-wavelength limits (Eq. (47), magenta lines).

Figs. 2-9 show that, for both the parallel plates and rectangular chamber models, the simulated (blue dashes) and theoretical (red circles) *average* LSC impedances match quite well for the cases $r_0$ = 0.5 cm, 1.0 cm and 1.5 cm ($r_0/h$≈0.21, 0.42, and 0.63). For the case of $r_0$=2.0 cm ($r_0/h$ ≈ 0.83), the relative errors between the theoretical and simulated peak LSC impedances are about 3.8% and 4.0% for the parallel plates and rectangular chamber models, respectively. This shows the line charge approximation in calculation of the image fields of a round beam is valid. Only at $r_0$ = 2.0 cm may this assumption underestimate the shielding effects of the image fields resulting in overestimation of the LSC impedances to some small noticeable extents. When the transverse beam dimension approaches the chamber height, the line charge assumption for the image charge fields of a round beam may induce bigger but still acceptable errors. For the wavelengths in the range of 0<$\lambda$≤5 cm, the theoretical (red circles) and simulated (blue dashes) *average* LSC impedance curves overlap the impedance curves for a beam in free space (black lines with circles) predicted by Eq. (46). It denotes that the shielding effects due to the image charges are on a negligible level, it is valid to calculate the average LSC impedances by Eq. (46) directly for the parallel plates and rectangular chamber models. For $\lambda$>5 cm, the average LSC impedances predicted by the model of a round beam in free space (black lines with circles) gradually deviate from and are larger than the theoretical (red circles) and simulated (blue dashes) LSC impedances of the two models discussed in this paper. This is caused by the neglect of the important shielding effects of beam chambers at large wavelengths. When $\lambda$ approaches 35 cm, the theoretical (red circles) and simulated (blue dashes) *average* LSC impedance curves approach the magenta curves predicted by Eq. (48) in Figs. 2-5 and Eq. (47) in Figs. 6-9 in the *long-wavelength* limits, respectively. These comparison results indicate the derived average LSC impedance formulae Eqs. (39)-(42) are consistent well with the simulations and the existing LSC impedance models in both the short-wavelength and long-wavelength limits. In the long-wavelength limits, for $r_0$<<$h$, the average LSC impedances of the round chamber model (green lines) are consistent with the ones predicted by the parallel plates and rectangular models (see the red circles and blue dashes in Fig. 2 and Fig. 6); while as $r_0$ increases and approaches $h$, the round chamber model (green lines) predicts smaller LSC impedances gradually than the parallel plates and rectangular chamber models (red circles and blue dashes) at large wavelengths (see Figs. 3-5 and Figs. 7-9). This result indicates, at large perturbation wavelengths, the round chamber model has larger shielding effects on the LSC fields than the models with planar and rectangular boundaries, and the shielding effects of the round chamber become more significant when $r_0/h$→1.

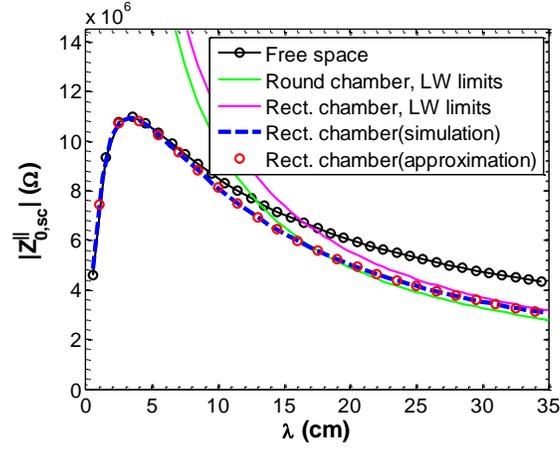

Fig. 6. Comparisons of the *average* LSC impedances of a *round* SIR beam with beam radius $r_0$=0.5 cm under different boundary conditions and in different wavelength limits. In the legend, 'Free space', 'Round chamber', and 'Rect. chamber' are boundary conditions, where 'Rect.' is the abbreviation for 'Rectangular'; The other symbols and abbreviations are the same as those in Fig. 2.

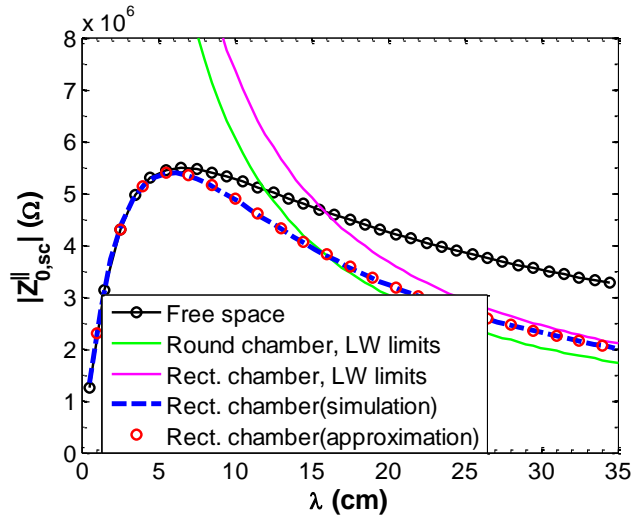

Fig. 7. Comparisons of the *average* LSC impedances of a *round* SIR beam with beam radius $r_0$=1.0 cm under different boundary conditions and in different wavelength limits.

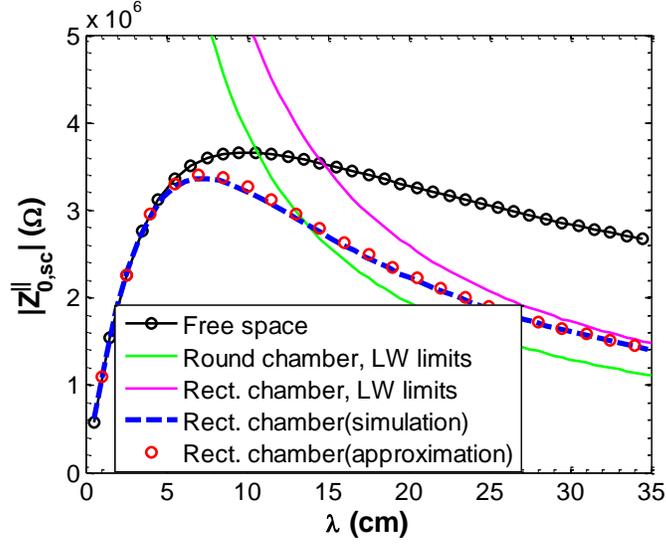

Fig. 8. Comparisons of the *average* LSC impedances of a *round* SIR beam with beam radius $r_0$=1.5 cm under different boundary conditions and in different wavelength limits.

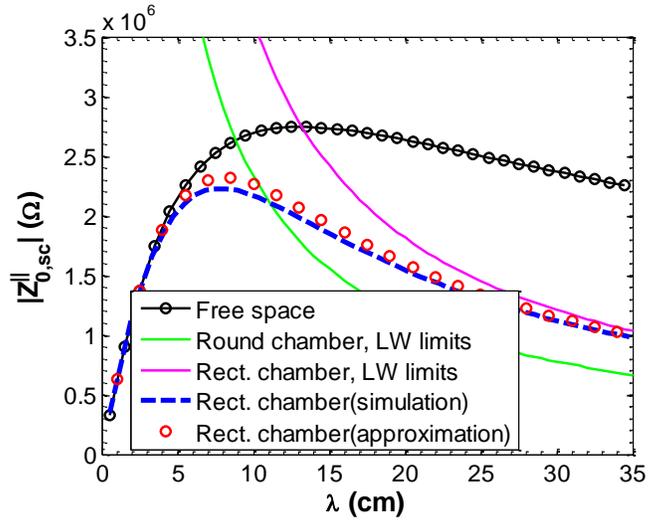

Fig. 9. Comparisons of the *average* LSC impedances of a *round* SIR beam with beam radius $r_0$ =2.0 cm under different boundary conditions and in different wavelength limits.

Fig. 10 shows the *average* LSC impedances of a SIR beam with beam radii $r_0$=0.5 cm and 2.0 cm midway between a pair of parallel plates with $h$=2.4 cm. The theoretical impedances are calculated by both the round beam model using Eq. (39) and square beam model ($a=b=r_0$) [17], respectively. For a round beam, the square beam model with length of side $a=b=r_0$ may underestimate the average LSC impedances compared with the round beam model with radius $r_0$.

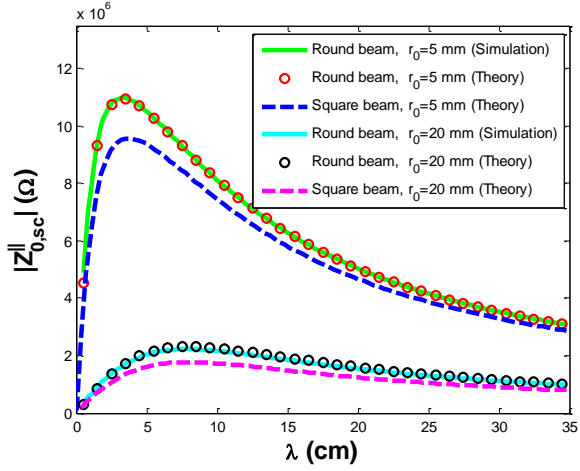

Fig. 10. Comparisons of the *average* LSC impedances between the round beam and square beam for a parallel plate field model. For a round beam, $r_0$ is the beam radius; for a square beam, $r_0$ is the half length of the side. The square beam model underestimates the LSC impedances.

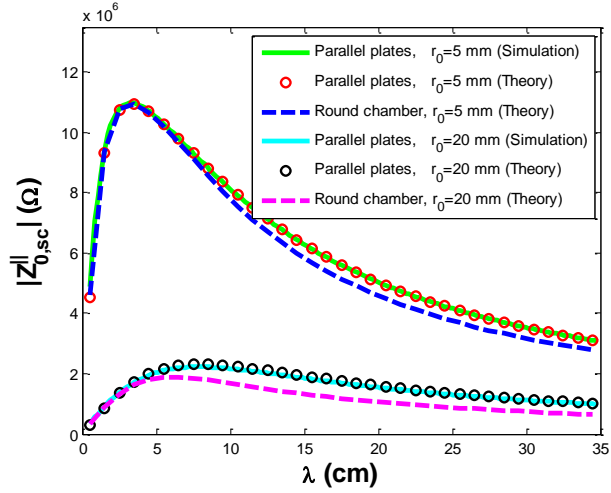

Fig. 11. Comparisons of the *average* LSC impedances of a round beam between parallel plates and a round beam inside a round chamber. The round chamber model underestimates the LSC impedances at larger $\lambda$.

Fig. 11 shows the *average* LSC impedances of a *round* SIR beam with beam radii $r_0$ =0.5 cm and 2.0 cm midway between a pair of *parallel plates* with $h$=2.4 cm and inside a *round chamber* with $r_w$=$h$=2.4 cm. The theoretical impedances of the parallel plates model and round chamber model are calculated by Eq. (39) and Eq. (43), respectively. For a round beam with fixed radius and energy, the round chamber model may underestimate the average LSC impedances compared with the parallel plates model with radius $h$=$r_w$. This difference is caused by the stronger shielding effects of the image fields produced by the round chamber compared with the parallel. Some literatures use the round chamber model to approximate the LSC field and impedance of a round beam between parallel plates or inside a rectangular chamber (e.g., Ref. (3)). Fig. 11 clearly indicates that this approximation only holds when $\lambda$ is small, where the shielding effect is negligible. For a 20 keV SIR beam with $r_0$ =0.5 cm inside a rectangular chamber with $w$=5.7 cm and $h$=2.4 cm, the round chamber approximation for the LSC impedance is only accurate for $\lambda \leq$ 5 cm. For $\lambda >$ 5 cm, the round chamber approximation will induce larger errors.

In summary, Figs. 2-11 show, for a typical SIR beam with 20 keV, $r_0$ =0.5 cm inside a rectangular chamber, when $\lambda \leq$ 5 cm, the image charge effects are negligible. In this case, for simplicity, we can use the LSC impedance formula for a beam in free space (Eq. (46)) to calculate the LSC impedance with a good accuracy; For $\lambda \geq$ 35 cm, we can use the impedance formulae in the long-wavelength limits Eq. (47) for a rectangular chamber model or Eq. (48) for a parallel plates model to estimate the LSC impedance. While for 5 cm<$\lambda$<35 cm, none of the existing models and formulae can be used to evaluate the LSC impedance accurately. In this case, we have to use the approximate theoretical impedance formulae Eqs. (41) and (42) for a rectangular chamber model or Eqs. (39) and (40) for a parallel plates model. This is the merit of the approximate analytical LSC impedance formulae derived in this paper.

## 5. Conclusions

In this paper, we mainly derive the approximate average LSC impedance formulae for a round beam under two boundary conditions: (a) Midway between a pair of infinitely large, perfectly conducting parallel plates (b) Inside and coaxial with a perfectly conducting rectangular chamber. In most accelerators, since $w>>r_0$, $h>>r_0$, the image charge fields of a round beam can be treated as those of a line charge in calculation of the LSC fields inside the beam. Consequently, the associated LSC impedances can be approximated by means of image methods based on the superposition theorem of the electric fields. The approximate theoretical average LSC impedances of the parallel plate model and the rectangular chamber model are consistent well with the numerical simulation results in a wide range of the radios of $r_0/h$. In addition, the theoretical LSC impedances predicted by the two field models also match well with the existing field models in both the short-wavelength ($\lambda \leq 5$ cm) and the long-wavelength ($\lambda \rightarrow 35$ cm) limits. In particular, for 5 cm<$\lambda$<35 cm, the approximate theoretical LSC impedances formulae have better accuracies than the existing models and formulae. Hence they are valid at any perturbation wavelengths and can be used as general expressions of the *average* LSC impedances in the future research work on space-charge induced instabilities, even for large ratios of $r_0/h$. At last, the image method together with line charge approximation employed in this paper can also be used to derive the LSC impedances of field models with other cross-sectional geometries.

## Acknowledgements

We would like to thank Prof. F. Marti and T. P. Wangler for their guidance. This work was supported by NSF Grant # PHY 0606007.